\documentclass[prl,twocolumn,superscriptaddress]{revtex4}
\usepackage{graphicx}

\begin{document}

\newcommand{\ket}[1]{\mbox{$|\!#1\;\!\rangle$}}
\newcommand{\aver}[1]{\mbox{$<\!#1\!\!>$}}
\def\ua{\uparrow}
\def\da{\downarrow}

\title{A cryogenic amplifier for fast real-time detection of single-electron tunneling}

\author{I. T. Vink}
\affiliation{Kavli Institute of Nanoscience, Delft University of
Technology,\\
PO Box 5046, 2600 GA Delft, The Netherlands}
\author{T. Nooitgedagt}
\affiliation{Kavli Institute of Nanoscience, Delft University of
Technology,\\
PO Box 5046, 2600 GA Delft, The Netherlands}
\author{R. N. Schouten}
\affiliation{Kavli Institute of Nanoscience, Delft University of
Technology,\\
PO Box 5046, 2600 GA Delft, The Netherlands}
\author{W. Wegscheider}
\affiliation{Institut f\"{u}r Angewandte und Experimentelle Physik,
Universit\"{a}t Regensburg, Regensburg, Germany}
\author{L. M. K. Vandersypen}
\affiliation{Kavli Institute of Nanoscience, Delft University of
Technology,\\
PO Box 5046, 2600 GA Delft, The Netherlands}

\date{\today}

\begin{abstract}
We employ a cryogenic High Electron Mobility Transistor (HEMT)
amplifier to increase the bandwidth of a charge detection setup with
a quantum point contact (QPC) charge sensor. The HEMT is operating
at 1K and the circuit has a bandwidth of 1 MHz. The noise
contribution of the HEMT at high frequencies is only a few times
higher than that of the QPC shot noise. We use this setup to monitor
single-electron tunneling to and from an adjacent quantum dot and we
measure fluctuations in the dot occupation as short as 400
nanoseconds, 20 times faster than in previous work.
\end{abstract}


\maketitle

The conventional method for studying quantum dot properties
electrically is to measure electron transport through the dot
\cite{Leo}. An alternative approach is to measure the current
through a quantum point contact (QPC) located next to the dot, which
is sensitive to the charge dynamics of the quantum dot
\cite{Lieven,Schleser,Gustavsson,Fujisawa,PettaCCO}. This technique
is very versatile and has also been used to probe the excited state
spectrum of a quantum dot \cite{JeroAPL,JohnsonChargeSensing},
perform single-shot read-out of electron spin states
\cite{JeroNature,RonaldPRL} and observe coherent electron spin
dynamics in quantum dots \cite{Petta}.

Until now, such a the current fluctuations through such a QPC charge
sensor has always been measured using a room temperature (RT)
current-to-voltage (IV-) convertor. This limits the measurement
bandwidth to several tens of kHz \cite{Lieven}, because of the
low-pass (LP) filter formed by the capacitance of the measurement
wires to ground and the input impedance of the amplifier. However,
increasing this bandwidth is crucial in order to study (real-time)
fast electron and nuclear spin dynamics \cite{HansonRMP} as well as
to increase the single-shot spin readout fidelity \cite{JeroNature}.
One way to increase the bandwidth is to embed the QPC in a resonant
circuit and measure its damping \cite{Reilly}, analogous to the 
operation of the RF-SET \cite{Schoelkopf}. In theory such an "RF-QPC" 
allows for single-shot charge detection within a few tens of nanoseconds 
\cite{SchoelkopfPrivate}. However, this technique requires
RF-modulation and is experimentally rather involved.

Here, we explore a much simpler approach to increasing the
bandwidth, which uses a HEMT operated in DC as a cryogenic
pre-amplifier \cite{CryoAmp}. Compared to a RT amplifier, a
cryogenic amplifier can be mounted much closer to the sample, which
significantly reduces the capacitance of the measurement wire. The
use of a HEMT has the additional advantage that the noise level at
cryogenic temperatures is very low (especially at high frequencies),
so a better charge sensitivity can be obtained.

\begin{figure}[b!]
\includegraphics[width=3.4in]{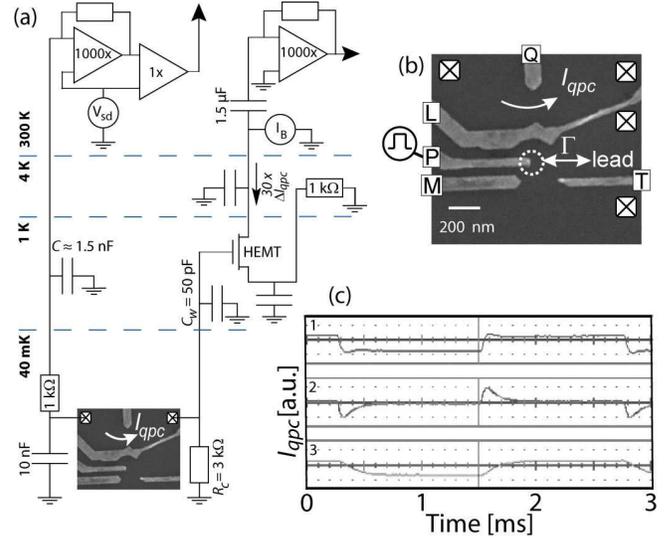}
\caption{(a) Schematic of the experimental setup. $R_{c}$ converts
fluctuations in $I_{qpc}$ into voltage fluctuations on the HEMT
gate. Through its transconductance the HEMT converts these
fluctuations into current fluctuations which are amplified by an
additional amplification stage at room temperature. $R_{c}$ and
$C_{w}$ form a 1 MHz LP-filter. (b) Scanning Electron Micrograph of
a similar device as used in the experiment. The dot (dashed circle)
and QPC are defined in a 2DEG formed at a GaAs/AlGaAs interface 90
nm below the surface, with an electron density of $1.3\times10^{15}$
m$^{2}$ by applying negative voltages to gates L,M, T and Q. Fast
voltage pulses can be applied to gate P. The crosses represent Ohmic
contacts. (c) Response to a voltage pulse applied to gate P. Trace 1
shows the total response to a voltage pulse when
$G_{qpc}\!\approx\!\frac{e^2}{h}$. When the QPC is pinched-off,
there is still a response due to crosstalk between the pulse line
and the HEMT gate-wire (trace 2), providing a measure for the
bandwidth of the readout circuit from the HEMT gate up to RT
($\sim8$ MHz). Subtracting trace 2 from 1 reveals the signal from
the QPC (trace 3) with a rise time of 285 ns, corresponding to a
bandwidth of 1 MHz.} \label{Setup}
\end{figure}

The HEMT is connected to the right lead of the QPC, which is also
connected to ground via  $R_{c}$ (Fig. \ref{Setup}(a)). A bias
voltage, $V_{sd}$, is applied to the left lead and a current
$I_{qpc}(t)$ will flow which depends on the QPC conductance
$G_{qpc}(t)$. The voltage over $R_{c}$ is a measure for this current
and is probed via the HEMT. Fluctuations of $G_{qpc}$ result in
fluctuations of $I_{qpc}$, denoted by $\Delta I_{qpc}$. These
generate voltage fluctuations on the HEMT gate with respect to the
voltage on its source, $V_{gs}$. The modulation of $V_{gs}$ results
in a modulation of the drain-source current, $I_{ds}$, through the
HEMT channel. This current is measured by an AC-coupled IV-convertor
at RT and digitized using a digital oscilloscope (LeCroy WaveRunner
6030A).

We use a commercially available HEMT (Agilent ATF 35143) with a 400
$\mu$m gate length and a threshold voltage $V_{t}\approx$ 0.4 V.
When appropriately biased (by controlling $I_{B}$), the
transconductance of the HEMT is $g_{m}=10$ mA/V, which relates the
drain-source current $I_{ds}$ through the HEMT to $V_{gs}$ as
$I_{ds}=-g_{m}V_{gs}$ implying $\Delta I_{ds}\approx-30 \Delta
I_{qpc}$, using $R_{c}=$ 3 k$\Omega$. The power dissipation of the
HEMT is 30 $\mu$W. In addition to the HEMT, $I_{qpc}$ can also be
measured \textit{simultaneously} in a 100 Hz bandwidth using a
IV-convertor at RT which is connected to the left lead of the QPC.
We refer to this measured current as the time averaged current.
\begin{figure}
\includegraphics[width=3.4in]{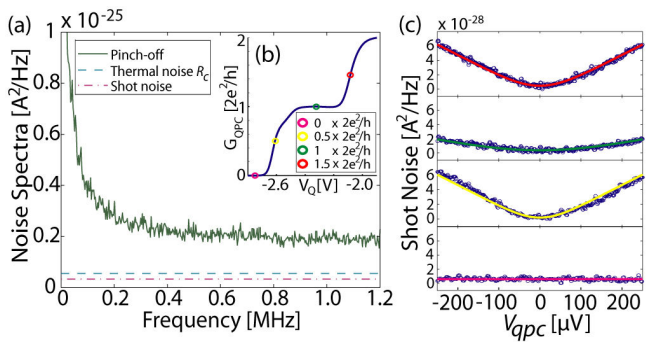}
\caption{(a) Noise spectrum of the setup including the cryogenic
HEMT amplifier. The measured spectrum is taken for the QPC in
pinch-off, thereby excluding shot noise and noise coming from the
other side of the QPC. The calculated noise contributions from the
QPC shot noise and the thermal noise of $R_{c}$ are plotted for
reference (dash-dotted and dashed line respectively). (b) QPC
conductance as a function of the voltage on gate Q. (c) Measurements
of the QPC shot noise power measured at the QPC conductances
indicated by the colored markers in (b). Solid lines are fits to Eq.
(\ref{Shotnoise}).} \label{NoiseFig}
\end{figure}
The quantum dot and the QPC are defined in a two-dimensional
electron gas (2DEG) by applying negative voltages to metal surface
gates (labeled L, M, T and Q in Fig \ref{Setup}(b). Gate L
completely separates the QPC source and drain electrically from the
leads of the dot. The experiment is performed in a dilution
refrigerator with a base temperature of 40 mK and with zero
externally applied magnetic field.

First, we characterize the bandwidth of the setup. The bandwidth
($BW$) is expected to be limited by the resistor $R_{c}$ and the
capacitance, $C_{w}$, of the measurement wire connecting the right
lead of the QPC to the HEMT gate ($BW=\left(2\pi
R_{c}C_{w}\right)^{-1}$). The HEMT is mounted on the 1K-stage, since
this has sufficient cooling power to dissipate the heat generated by
the HEMT in operation. The value for $C_{w}$ is then a tradeoff
between two requirements: a low capacitance and sufficient thermal
anchoring of the wire. The value of $R_{c}$ is also a tradeoff:
increasing the value of $R_{c}$ increases the amplitude of the
voltage fluctuations on the HEMT gate ($\Delta V_{gs}=\Delta I_{qpc}
R_{c}$) but reduces the bandwidth of the setup (for a given value of
$C_{w}$). Our aim is to detect single-electron tunneling on a
sub-microsecond timescale. The value for $R_{c}$ was chosen assuming
$\Delta I_{qpc}\approx$ 400 pA and an input referred voltage noise
0.4 nV/$\sqrt{\textrm{Hz}}$. $R_{c}=$ 3 k$\Omega$ then gives SNR
$\approx$ 3 and a bandwidth of 1 MHz. The bandwidth is determined by
measuring the QPC response to fast voltage pulses applied to gate P.
The measured rise times are 285 ns, yielding a bandwidth of 1 MHz,
in excellent agreement with the designed bandwidth (Fig.
\ref{Setup}(c)).

The next step is a characterization of the noise level. We measure
the total noise spectral density and plot this as an input referred
current noise in Fig. \ref{NoiseFig}(a). A characteristic $1/f$
contribution is present up to 200 kHz. For frequencies above 200
kHz, the spectrum is approximately flat, saturating at
$0.2\times10^{-25} \mathrm{A}^{2}/\mathrm{Hz}$ (= 0.4
nV/$\sqrt{\mathrm{Hz}}$). This is very close to the voltage
fluctuations generated by the QPC shot noise (calculated to be
$S_{I}=0.17$ nV/$\sqrt{\mathrm{Hz}}$, for 1 mV bias over the QPC
\cite{Lieven}). We test this by a \textit{direct} measurement of the
QPC shot noise. We measure the rms voltage after band-pass filtering
the output of the RT IV-convertor (bandwidth from 500 kHz to 1 MHz).
In Fig. \ref{NoiseFig}(b) we show the QPC conductance $G_{qpc}$ as a
function of the voltage on gate Q, determined from the time averaged
current. The colored markers indicate the QPC conductances
($G_{qpc}=n\frac{e^2}{h}$, $n=0,1,2,3$) at which the shot noise was
measured as a function of bias over the QPC, $V_{qpc}$, see Fig.
\ref{NoiseFig}(c). $V_{qpc}$ is varied by changing $V_{sd}$. We
verified that the QPC was in its linear regime for the entire range
of $V_{qpc}$. The shot noise spectral density $S_{I}$ can be
expressed as \cite{Blanter,DiCarlo}
\begin{equation}\label{Shotnoise}
S_{I}=\frac{2e^2}{h}\sum_{i}\mathcal{N}_{i}\left[eV_{qpc}\coth\left(\frac{eV_{qpc}}{2k_{B}T_{e}}\right)-2k_{B}T_{e}\right]
\end{equation}
where $\mathcal{N}_{i}=T_{i}\left(1-T_{i}\right)$ with $T_{i}$ the
QPC transmission coefficient of mode $i$, $V_{qpc}$ the bias over
the QPC, $k_{B}$ the Boltzmann constant and $T_{e}$ the electron
temperature. The solid lines in Fig. \ref{NoiseFig}(c) are fits to
Eq. (\ref{Shotnoise}) yielding $\mathcal{N}=0.234, 0.090, 0.229$ and
$0$ from top to bottom, in agreement with the QPC conductances. The
measurements prove that the input referred voltage noise is indeed
very close to the shot noise limit in this setup. From the fits we
also extract the electron temperature $T_{e}=255$ mK, consistent
with the value
obtained from the width of Coulomb peaks ($T_{e}=267$ mK).\\
The noise measurements show that the noise from the HEMT is in
agreement with our initial estimation. We therefore expect to have
sufficient SNR to detect single-electron tunnel events. To test this
experimentally, the dot is tuned to be near the $0\leftrightarrow 1$
electron transition by adjusting the voltages on gates L, M and T,
and to be isolated from the bottom lead \cite{JeroAPL}. The dot
remains coupled to the other lead with a tunable tunnel rate,
$\Gamma$. An electron is now allowed to tunnel back and forth
between the dot and the lead and the QPC current should therefore
exhibit a random telegraph signal (RTS). The QPC conductance is set
again at approximately $\frac{e^2}{h}$.
\begin{figure}
\includegraphics[width=3.4in]{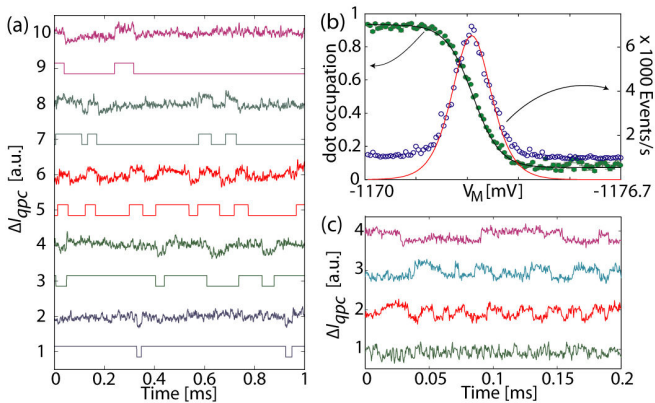}
\caption{(a) Measured QPC current when increasing the dot potential
from top to bottom. The result of our flank detection routine is
plotted below each measured trace. An additional band-pass filter
(200 Hz - 200 kHz) was used for this measurement. (b) Dot occupation
extracted from the same data as (a) as a function of $V_{M}$. From
the same data we extract the number of tunnel events per second as a
function of $V_{M}$ from which we can extract the tunnel rate
\cite{TraceLengthSaturation}. The solid curves are fits to the data
(see text). (c) The tunnel rate $\Gamma$ is increased from top to
bottom by decreasing the negative voltage on gate T. Here, the
signal was band-pass filtered from 3 kHz to 1 MHz. The shortest
detectable events are on the order of 400 ns.} \label{Fig3}
\end{figure}
In order to maximize $\Delta I_{qpc}$, we want to apply the highest
possible bias, $V_{qpc}$. However, for $V_{qpc} > 0.65$ mV, we
observe a severe change in the dot occupation, most probably due to
intradot excitations to the first orbital excited state
\cite{Eugen}. We therefore restrict ourselves to QPC bias voltages
below 0.65 mV. This reduces $\Delta I_{qpc}$ to 320 pA, resulting in
a lower SNR. Measurements of the RTS are shown in Fig. \ref{Fig3}.
To verify that the measured RTS originates from electron tunnel
events between the dot and the lead, we varied two control
parameters, as in \cite{Lieven}: (1) the dot electrochemical
potential $\mu$ relative to the Fermi level of the lead $\mu_{F}$
and (2) the tunnel barrier between the dot and the lead. The dot
potential is changed by changing the voltage on gate M. The dot
occupation probability $P$ depends on $\mu-\mu_{F}$ and the
temperature broadening of the lead so it should directly reflect the
Fermi-Dirac distribution of electronic states in the lead. We infer
the dot occupation from the measured average time the electron
spends on (off) the dot, $\tau_{on(off)}$, as
$P=\frac{\tau_{off}}{\tau_{on}+\tau_{off}}$, \cite{Schleser}.
However, since both the HEMT and the RT IV-convertor AC-coupled,
signals from the QPC are high-pass filtered (1.2 kHz cut-off). We
can therefore not use a simple threshold detection scheme
\cite{JeroNature} but instead detect the flanks of the steps in
$\Delta I_{qpc}$ to obtain the single-electron tunneling statistics.
In Fig. \ref{Fig3}(b) the average dot occupation is plotted versus
the voltage on gate M ($V_{M}$). At $V_{M}$= - 1172.8 mV, $\mu$ is
aligned with $\mu_{F}$. The solid black line is a fit to the
Fermi-Dirac distribution function $f(\mu)$ yielding an electron
temperature $T_{e}$ = 275 mK. The average times $\tau_{on/off}$ also
allow the determination of the tunnel rate $\Gamma$. The Fermi
distribution and the tunnel rate $\Gamma$ determine the average
number of tunnel events per second as
$r_{e}=1/(\tau_{on}+\tau_{off})=\Gamma\times
f(\mu)\left[1-f(\mu)\right]$. This is also plotted in Fig.
\ref{Fig3}(b). The fit to this data yields $\Gamma$ = 26.1 kHz
(solid red line) \cite{TraceLengthSaturation}. The tunnel rate
$\Gamma$ can be varied via the voltage on gate T (Fig. \ref{Fig3}
(c)). The shortest detectable events are on the order of 400 ns. The
charge sensitivity reached is $4.4 \times 10^{-4}
e/\sqrt{\mbox{Hz}}$ in the range 200 kHz - 1 MHz, only 3.8 times
larger than the shot noise limit in this setup with $V_{qpc}=$ 0.65
mV.

We have demonstrated that a HEMT can be used as a cryogenic
amplifier to increase the measurement bandwidth of a QPC charge
detection setup. The bandwidth of the setup is 1 MHz and the input
referred voltage noise is measured to be 0.4 nV/$\sqrt{\mathrm{Hz}}$
above $\sim$200 kHz, which is close to the QPC shot noise limit.
This allows us to detect fluctuations in the dot occupation as short
as 400 ns, 20 times faster than previously achieved using a QPC as a
charge sensor. The bandwidth could be further increased by placing
the HEMT even closer to the sample (since the dissipation in the
HEMT is low enough), which would reduce the capacitance even more. A
lower amplifier noise (both $1/f$ and baseline) could be obtained by
using a HEMT with a larger gate area.

We thank F.H.L. Koppens, L.P. Kouwenhoven, J. Love, T. Meunier, K.C.
Nowack, J.H. Plantenberg, R.J. Schoelkopf, G.A. Steele, H.P. Tranitz
and L.H. Willems van Beveren for help and discussions and A. van der
Enden and R.G. Roeleveld for technical support. This work was
supported by the Dutch Science Foundation (FOM and NWO).


\begin{references}
\bibitem{TraceLengthSaturation}
For every trace without a tunnel event, the total length of the
trace (1 ms) is assigned to $\tau_{on/off}$. This results in a
baseline of 1000 events/s instead of 0. To determine $\Gamma$, we
fit to a subset of the data with a sufficient number of
tunnel events.
\end{references}
\end{document}